\documentclass[twocolumn]{aastex631}

\usepackage{subcaption}
\usepackage{threeparttable} 
\usepackage{amsmath}

\begin{document}

\title{Building on the archives: Connecting the CN/CO intensity ratio with global galaxy properties in nearby U/LIRGs}

\author[0000-0003-4577-8644]{Blake Ledger}
\affiliation{Department of Physics \& Astronomy, University of Victoria, Finnerty Road, Victoria, British Columbia, V8P 1A1, Canada}

\author[0000-0001-5817-0991]{Christine D. Wilson}
\affiliation{Department of Physics and Astronomy, McMaster University, 1280 Main St. W., Hamilton, Ontario L8S 4M1, Canada}

\author{Osvald Klimi}
\affiliation{Department of Physics and Astronomy, McMaster University, 1280 Main St. W., Hamilton, Ontario L8S 4M1, Canada}

\author[0000-0003-3638-8943]{N\'{u}ria Torres-Alb\`{a}}
\altaffiliation{GECO Fellow}
\affiliation{Department of Physics and Astronomy, Clemson University, Kinard Lab of Physics, Clemson, SC 29634, USA}
\affiliation{Department of Astronomy, University of Virginia, P.O. Box 400325, Charlottesville, VA 22904, USA}

\author[0000-0002-2501-9328]{Toshiki Saito}
\affiliation{Faculty of Global Interdisciplinary Science and Innovation, Shizuoka University, 836 Ohya, Suruga-ku, Shizuoka 422-8529, Japan}




\begin{abstract}
We use the CN/CO intensity ratio to obtain the dense gas fraction, $f_{\text{dense}}$, for a sample of 16 Ultraluminous and Luminous Infrared Galaxies and compare $f_{\text{dense}}$ with a suite of global galaxy properties. We find a significant correlation between $f_{\text{dense}}$ and star formation rate calculated using both infrared luminosities and radio continuum, although there is significant scatter in each relation. We find no trend between global or peak $f_{\text{dense}}$ and merger stage. We find no correlation between global $f_{\text{dense}}$ and X-ray luminosity; however, the correlation becomes significant when we measure $f_{\text{dense}}$ at the location of peak X-ray emission. Our interpretation is that the dense gas is co-localized with strong X-ray emission from an active galactic nuclei or strong central star formation.

\end{abstract}



\section{Introduction}
\label{sec:intro}

Dense molecular gas plays a crucial role in connecting star formation and the interstellar medium (ISM) in galaxies. The dense phase of molecular gas is usually traced using hydrogen cyanide (HCN; \citealt{Gao2004a}), while the bulk molecular gas is traced using carbon monoxide (CO; \citealt{Bolatto2013}). The dense gas fraction ($f_{\text{dense}}$) in galaxies is typically estimated using the HCN/CO intensity ratio (e.g., \citealt{Gao2004a, Garcia2012}). $f_{\text{dense}}$ has been found to anti-correlate with the star formation efficiency (e.g., \citealt{Usero2015, Jimenez2019}). Some studies suggest that this anti-correlation results from the conditions of the local environment (gas and stellar surface densities) and pressure in the molecular gas \citep{Gallagher2018, Jimenez2019, Neumann2023}.

Recently, we have proposed that the cyanide radical (CN) can be used to trace the dense molecular gas in a similar way to HCN \citep{Wilson2023}. \cite{Wilson2023} found a nearly constant CN/HCN intensity ratio in a sample of 9 nearby, star-forming galaxies on sub-kpc scales. The CN/HCN ratio showed no correlation with molecular gas surface density and only a slight trend with star formation rate surface density. Additionally, \cite{Wilson2023} found that the CN/CO and HCN/CO intensity ratios track each other well over at least a factor of 10. HCN is typically found to be optically thick in extragalactic studies (e.g., \citealt{Jimenez2017}), while CN is more commonly optically thin (e.g., \citealt{Tang2019, Ledger2024}). \cite{Wilson2023} suggest that the optically thick HCN line traces a population of gravitationally bound or collapsing subclumps within molecular clouds indirectly, while the optically thin CN line traces the same dense gas directly. In this paper, we will use the CN/CO intensity ratio as an estimator of dense gas fraction.

With their large gas reservoirs, high dense gas fractions, and high star formation rates (SFRs), U/LIRGs are ideal laboratories in which to study molecular gas and star formation \citep{Sanders1996, Lonsdale2006, Perez2021}. They often host active galactic nuclei (AGN), with $40-50\%$ of U/LIRGs having a Seyfert nucleus and strong AGN \citep{Veilleux1995, Sanders1996, Iwasawa2011}. U/LIRGs can have significant hard X-ray emission from the AGN region, high-mass X-ray binaries (HMXBs), and other star formation activities \citep{Pereira2011, Iwasawa2011, Torres2018}. X-rays can alter the physical and chemical properties of the ISM (e.g., \citealt{Meijerink2005, Meijerink2007, Viti2014, Kawamuro2021}) via negative AGN feedback and/or molecular gas destruction \citep{Kawamuro2019, Kawamuro2020, Kawamuro2021}. On the other hand, dense molecular gas is often associated with circumnuclear disks (CNDs) around AGN that supply gas for accretion onto central supermassive black holes (e.g., \citealt{Izumi2016}).

\cite{Ledger2024}, hereafter ``Paper I'', measured the CN/CO intensity ratio in a sample of 16 U/LIRGs. In this paper, we convert these intensity ratios to dense gas fractions and compare $f_{\text{dense}}$ to global galaxy properties, such as SFR, merger stage, and hard X-ray luminosity. Section \ref{sec:data} describes our sample, data analysis, and the conversion of our measured quantities to physical properties (e.g., $f_{\text{dense}}$ and SFR). Section \ref{sec:results} describes the correlations (or lack thereof) between $f_{\text{dense}}$ and global galaxy properties, as well as our physical interpretation of the correlations. We summarize our conclusions in Section \ref{sec:conc}.

\begin{figure*}
    \includegraphics[width=0.99\textwidth]{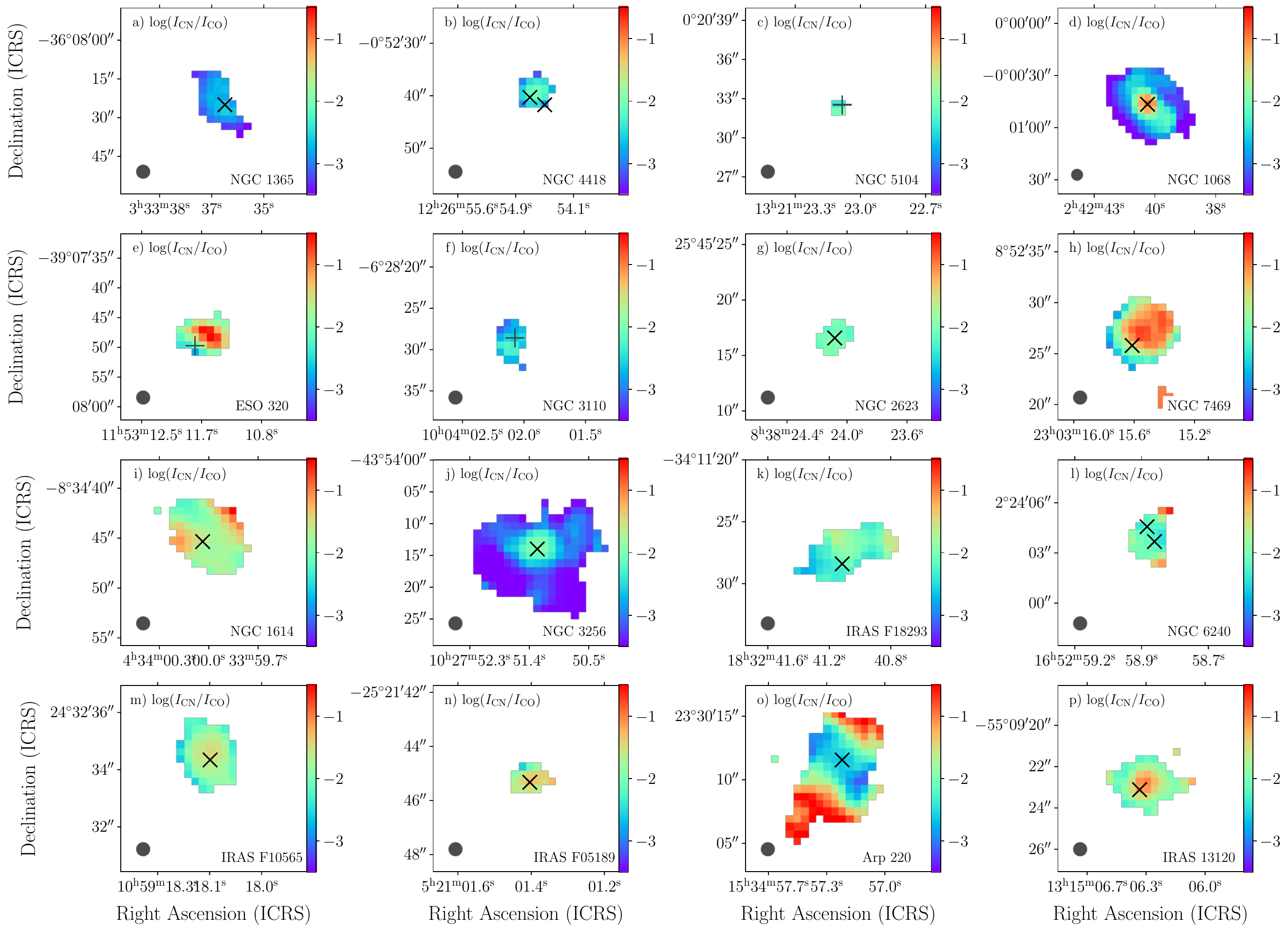}
    \caption{Log-scale $I_{\text{CN}}/I_{\text{CO}}$ ratio maps (K km s$^{-1}$) for the 16 galaxies in our sample, ordered by increasing infrared luminosity. The maps have $250$ pc pixels and the individual beams, which have been smoothed to a $500$ pc diameter, are shown for scale in the bottom left corner for each galaxy. Pixels shown here are detected in CO and both CN hyperfine groupings, but the measured ratio includes only the brightest CN hyperfine grouping \citep{Ledger2024}. Black crosses show the position of peak hard X-ray emission from \textit{Chandra} observations, where NGC 4418 and NGC 6240 have two peaks. NGC 5104, ESO 320-G030, and NGC 3110, which were not observed by \textit{Chandra}, have gray plus signs at the position of peak CO emission. In most galaxies, the hard X-ray peak roughly matches the peak CO and radio continuum emission.}
    \label{fig:CN_CO_maps_w_xray}
\end{figure*}

\begin{table*}
\centering
    \caption{Dense gas and star formation properties for the U/LIRG sample.}%
    \begin{tabular}{c|cc|cc|cl|cc}
    \hline
        Galaxy\textsuperscript{\textit{a}} & $I_{\text{CN}}/I_{\text{CO}}$\textsuperscript{\textit{b}} & $f_{\text{dense}}$ & $I_{\text{CN}}/I_{\text{CO}}$ \textsuperscript{\textit{c}} & $f_{\text{dense}}$ & $L_{\text{IR}}$\textsuperscript{\textit{d}} & SFR$_\text{IR}$\textsuperscript{\textit{e}} & $S_{\text{110 GHz}}$\textsuperscript{\textit{f}} & SFR$_\text{110 GHz}$\textsuperscript{\textit{e}} \\
        & (global) & (global) & (at peak) & (at peak) & ($L_{\odot}$) & ($M_{\odot}$ yr$^{-1}$) & (mJy) & ($M_{\odot}$ yr$^{-1}$)  \\
        \hline
        IRAS 13120-5453     & $0.17$ & $0.61$ & $0.31$ & $1.12$ & $12.28$ & $283.6$, $\ast$ & $39.6 \pm 2.2$ & $634.0$ \\
        Arp 220             & $0.09$ & $0.32$ & $0.07$ & $0.25$ & $12.21$ & $241.4$, $\ast$ & $46.6 \pm 10.9$ & $265.2$ \\
        IRAS F05189-2524    & $0.10$ & $0.36$ & $0.27$ & $0.97$ & $12.16$ & $215.2$, $\ast$ & $2.5 \pm 0.8$  & $73.9$ \\
        IRAS F10565+2448    & $0.08$ & $0.29$ & $0.22$ & $0.79$ & $12.07$ & $174.9$ & $3.1 \pm 0.5$  & $98.4$ \\
        NGC 6240            & $0.08$ & $0.29$ & $0.11$ & $0.40$ & $11.85$ & $105.4$, $\ast$ & $11.3 \pm 2.3$ & $109.5$ \\
        IRAS F18293-3413    & $0.05$ & $0.18$ & $0.11$ & $0.40$ & $11.79$ & $91.8$ & $18.9 \pm 1.5$  & $97.3$ \\
        NGC 3256            & $0.02$ & $0.07$ & $0.14$ & $0.50$ & $11.75$ & $83.7$, $\ast$ & $38.2 \pm 1.8$  & $65.4$ \\
        NGC 1614            & $0.15$ & $0.54$ & $0.20$ & $0.72$ & $11.65$ & $66.5$, $\ast$ & $11.5 \pm 2.3$  & $46.1$ \\
        NGC 7469            & $0.10$ & $0.36$ & $0.22$ & $0.79$ & $11.59$ & $57.9$, $\ast$ & $19.1 \pm 1.4$  & $72.2$ \\
        NGC 2623            & $0.07$ & $0.25$ & $0.13$ & $0.47$ & $11.59$ & $57.9$, $\ast$ & $7.8 \pm 2.1$   & $46.7$ \\
        NGC 3110            & $0.02$ & $0.07$ & $0.07$ & $0.25$ & $11.35$ & $33.3$ & $6.3 \pm 0.5$   & $32.7$ \\
        ESO 320-G030        & $0.11$ & $0.40$ & $0.12$ & $0.43$ & $11.35$ & $33.3$ & $12.0 \pm 2.7$  & $26.9$ \\
        NGC 1068            & $0.06$ & $0.22$ & $0.39$ & $1.40$ & $11.29$ & $29.0$, $\ast$ & $129.5 \pm 31.6$ & $22.2$ \\
        NGC 5104            & $0.07$ & $0.25$ & $0.09$ & $0.32$ & $11.21$ & $24.1$ & $3.3 \pm 0.5$   & $20.6$ \\
        NGC 4418            & $0.08$ & $0.29$ & $0.12$ & $0.43$ & $11.16$ & $21.5$, $\ast$ & $11.8 \pm 4.0$  & $12.8$ \\
        NGC 1365            & $0.03$ & $0.11$ & $0.07$ & $0.25$ & $11.08$ & $17.9$, $\ast$ & $31.5 \pm 3.7$  & $10.6$ \\
    \hline
    \end{tabular}
    \label{tab:measured_data}
    \begin{tablenotes}
        \item \textbf{Notes:} \textsuperscript{\textit{a}}Ordered by decreasing infrared luminosity.
        \item \textsuperscript{\textit{b}}Global $I_{\text{CN}}/I_{\text{CO}}$ ratios as measured in Paper I. Measurement uncertainties for the global $I_{\text{CN}}/I_{\text{CO}}$ ratios include a 5\% ALMA calibration uncertainty and are all $\pm0.01$, except NGC 5104, which is $\pm0.02$ because of the small number of detected pixels (Figure \ref{fig:CN_CO_maps_w_xray}).
        \item \textsuperscript{\textit{c}}$I_{\text{CN}}/I_{\text{CO}}$ at the location of the peak hard X-ray emission from \textit{Chandra}. For the three galaxies with no \textit{Chandra} data (NGC 5104, ESO 320-G030, NGC 3110), we give $I_{\text{CN}}/I_{\text{CO}}$ at the peak CO intensity. For the two galaxies with double X-ray peaks (NGC 4418, NGC 6240), we use the peak X-ray pixel which is closest to the location of the peak CO intensity. We estimate uncertainties of $\pm10\%$ for $I_{\text{CN}}/I_{\text{CO}}$ at peak because of using only a single pixel.
        \item \textsuperscript{\textit{d}}Infrared luminosities from GOALS \citep{Armus2009} scaled to the distances given in Table \ref{tab:sample_data}.
        \item \textsuperscript{\textit{e}}Galaxies with an AGN (see Table \ref{tab:sample_data}) have been identified with an $\ast$ to indicate which IR-based SFRs may suffer from AGN contamination.
        \item \textsuperscript{\textit{f}}Uncertainties on the radio continuum flux density were calculated as $\sigma_{\text{RMS}} \times \sqrt{N_{\text{beam}}}$, where $\sigma_{\text{RMS}}$ is the RMS noise of the continuum map in Jy beam$^{-1}$ and $N_{\text{beam}}$ is the number of beams in the aperture used to measure $S_{\text{110 GHz}}$. The 5\% ALMA calibration uncertainty is not included here. SFR$_{\text{110 GHz}}$ is not corrected for any dust contamination, synchrotron emission, or AGN impact (see Section \ref{subsection:sfr}).
    \end{tablenotes}
\end{table*}

\begin{table*}
\centering
    \caption{Global galaxy properties for the U/LIRG sample.}%
    \begin{tabular}{c|cc|cc|cc}
    \hline
        Galaxy\textsuperscript{\textit{a}} & Distance\textsuperscript{\textit{b}} & Merger stage\textsuperscript{\textit{c}} & $L_{\text{2-10 keV}}$\textsuperscript{\textit{d}} & Reference\textsuperscript{\textit{d}} & AGN$_{\text{MIR}}$\textsuperscript{\textit{e}} & Type of AGN\textsuperscript{\textit{f}} \\
         & (Mpc) & & ($10^{41}$ erg s$^{-1}$) & & \\        
        \hline
        IRAS 13120-5453     & 137   & d & $4.5$ & (1) & $0.2$ & Obscured \\
        Arp 220             & 81.1  & d & $0.68$ & (1) & $0.29$ & Obscured \\
        IRAS F05189-2524    & 188   & d & $130.0$ & (1) & $0.67$ & Seyfert 2 \\
        IRAS F10565+2448    & 194   & d & $1.6$ & (1) & $0.18$ & None \\
        NGC 6240            & 106   & d & $21.0$ & (1) & $0.29$ & Obscured \\
        IRAS F18293-3413    & 77.2  & c & $0.94$ & (1) & $0.17$ & None \\
        NGC 3256            & 44.3  & d & $0.921$ & (2) & $0.16$ & Obscured \\
        NGC 1614            & 67.9  & d & $1.236$ & (2) & $0.22$ & LINER \\
        NGC 7469            & 66    & a & $148.0$ & (4) & $0.44$ & Seyfert 1 \\
        NGC 2623            & 83.4  & d & $1.285$ & (2) & $0.26$ & LINER \\
        NGC 3110            & 77.8  & a & $0.87$ & (2) & $0.23$ & None \\
        ESO 320-G030        & 50.7  & N & $0.11$ & (3) & $0.17$ & None \\
        NGC 1068            & 13.97 & N & $1.015$ & (2) & $1.0$ & Seyfert 2 \\
        NGC 5104            & 84.6  & a & $0.15$ & (5) & $0.27$ & None \\
        NGC 4418            & 35.3  & N & $0.019$ & (2) & $0.62$ & Obscured \\
        NGC 1365            & 19.57 & N & $2.889$ & (2) & $0.51$ & Obscured \\
    \hline
    \end{tabular}
    \label{tab:sample_data}
    \begin{tablenotes}
        \item \textbf{Notes:} \textsuperscript{\textit{a}}Ordered by decreasing infrared luminosity.
        \item \textsuperscript{\textit{b}}Luminosity distances as compiled in Paper I.
        \item \textsuperscript{\textit{c}}Merger stages from \cite{Stierwalt2013}, where ``N'' is a non-merger, ``a'' is a pre-merger, ``c'' is a mid-stage merger, and ``d'' is a late-stage merger.
        \item \textsuperscript{\textit{d}}Hard X-ray luminosities from (1) C-GOALS I, \citealt{Iwasawa2011}, (2) C-GOALS II, \citealt{Torres2018}, (3) \citealt{Pereira2011}; (4) \citealt{Liu2014}; (5) \citealt{Privon2020}. The $L_{\text{2-10 keV}}$ values have been corrected for Galactic absorption in the original publications and scaled to the distances given here.
        \item \textsuperscript{\textit{e}}AGN$_{\text{MIR}}$ from \cite{DiazSantos2017}. AGN$_{\text{MIR}} >0.2$ indicates a galaxy with an AGN that contributes significantly to the bolometric luminosity, while AGN$_{\text{MIR}} <0.2$ indicates a galaxy where the energetics are heavily dominated by star formation. See \cite{DiazSantos2017} for more details on the determination of AGN$_{\text{MIR}}$ through the analysis of \textit{Spitzer} spectra.
        \item \textsuperscript{\textit{f}}From Paper I, where a literature search was used to classify AGN as obscured, optically identified (Seyfert 1, Seyfert 2, or LINER), or with no evidence of an AGN. 
    \end{tablenotes}
\end{table*}

\section{Data, analysis, and galaxy properties}
\label{sec:data}

\subsection{Galaxy sample}
\label{subsec:galaxies}

Paper I describes the selection process for our sample of 16 U/LIRGs. Briefly, the HERschel ULIRG Survey (HERUS; \citealt{Pearson2016}) and the Great Observatories All-Sky LIRG Survey (GOALS; \citealt{Armus2009}) were used as parent samples to identify nearby ($z<0.05$) U/LIRGs that have both the CO $J=1-0$ and CN $N=1-0$ lines detected in the Atacama Large Millimeter/Submillimeter Array (ALMA) archive at $\theta < 500$ pc resolution. 11 of the 16 U/LIRGs in the sample have known AGN, and there are 4 ULIRGs and 12 LIRGs. The measured properties of the galaxies in our sample are listed in Tables \ref{tab:measured_data} and \ref{tab:sample_data}, in order of decreasing infrared luminosity. For all our analysis, we use Ned Wright's (Updated) Cosmology Calculator adopting WMAP 5-year cosmology with $H_{0} = 70.5$ km s$^{-1}$ Mpc$^{-1}$, $\Omega = 1$, $\Omega_{\rm m} = 0.27$ in the 3K CMB frame.

\subsection{ALMA data analysis}
\label{subsec:alma}

\subsubsection{CO and CN lines}
\label{subsubsec:ALMA_CO_CN}

In this work, we consider CO ($J=1-0$), hereafter ``CO'', and the brightest hyperfine grouping of the CN ($N=1-0$) line, hereafter CN.\footnote{This hyperfine grouping of CN has a rest frequency of $\sim 113.49$ GHz and quantum number CN ($N = 1 - 0, J = 3/2 - 1/2$). Paper I discusses the fainter hyperfine grouping, CN ($N = 1 - 0, J = 1/2 - 1/2$), and its usefulness for determining CN optical depth.} The ALMA data for the 16 galaxies span 20 unique ALMA project codes from observing Cycles 1-6. We refer the reader to Paper I for specific details of the steps taken to process the data systematically to ensure uniform analysis. For each galaxy, Paper I provides integrated intensity maps for CO and CN (in units of K km s$^{-1}$), a ratio map including only detected pixels in CO and both CN hyperfine groupings, and integrated spectra. In Figure \ref{fig:CN_CO_maps_w_xray}, we compile the $I_{\text{CN}}/I_{\text{CO}}$ ratio maps for all 16 galaxies and adjust them to the same colour scale. The maps show the CN bright hyperfine grouping compared to CO, but include pixels detected only in CO and both CN hyperfine groupings. For the analysis in this paper, we consider the CO and CN cubes and the measured $I_{\text{CN}}/I_{\text{CO}}$ ratios from Paper I. For a comparison with single-dish $I_{\text{CN}}/I_{\text{CO}}$ global ratios, we refer the reader to Paper I or to \cite{Wilson2018}.

\subsubsection{Radio continuum}
\label{subsubsec:ALMA_radio}

We produced new radio continuum maps for 15 of our galaxies using Common Astronomy Software Applications (\textsc{casa}, \citealt{McMullin2007}) and the Physics at High Angular resolution in Nearby GalaxieS (PHANGS)-ALMA pipeline v3 \citep{Leroy2021} with \textsc{casa} version 5.6.3. The continuum image for NGC 1068 is from \cite{Saito2022a, Saito2022b}. The continuum at a frequency of $\sim110$ GHz was extracted from the calibrated ALMA data used in Paper I, masking out the spectral regions containing the CN and CO lines. Continuum emission at these frequencies is thought to be associated with free-free emission from recent star formation and is a useful tool for measuring the unobscured SFR (Section \ref{subsection:sfr}) in dusty galaxies like U/LIRGs (see e.g., \citealt{Wilson2019}). We smoothed the continuum images to a physical resolution of $500$ pc and rebinned and regridded the data to match the CN and CO images with Nyquist sampled pixel sizes of $250$ pc. Finally, the continuum maps were corrected by the primary beam. The global $110$ GHz radio flux density and its uncertainty (Table \ref{tab:measured_data}) were measured within an aperture encompassing the emission seen in the integrated intensity.

\subsection{Dense gas fractions}
\label{subsection:f_dense_calc}

The dense gas fraction is typically calculated as
\begin{equation}
    f_{\text{dense}} = \Sigma_{\text{dense}} / \Sigma_{\text{mol}},
    \label{eqn:fdense}
\end{equation}
where $\Sigma_{\text{dense}}$ is the surface density of dense gas and $\Sigma_{\text{mol}}$ is the surface density of molecular gas, both in units of $M_{\odot}$ pc$^{-2}$. $\Sigma_{\text{mol}}$ is typically obtained from the CO ($1-0$) intensity and a conversion factor, $\alpha_{\text{CO}}$ \citep{Bolatto2013}. In disk galaxies, like the Milky Way, the canonical value is $\alpha_{\text{CO}} = 4.35$, which includes a factor of 1.36 for Helium \citep{Bolatto2013}. In galaxies with high gas surface densities, like U/LIRGs, $\alpha_{\text{CO}}$ tends to decrease \citep{Downes1993, Solomon1992, Papadopoulos2012}. The canonical value (including Helium) of $\alpha_{\text{CO}} = 1.1$ \citep{Bolatto2013} has recently been confirmed observationally by \cite{He2024}.

Similarly, $\Sigma_{\text{dense}}$ is obtained from the HCN ($1-0$) intensity and a conversion factor, $\alpha_{\text{HCN}}$. \cite{Gao2004a} use a value of $\alpha_{\text{HCN}} = 10$, which becomes $\alpha_{\text{HCN}} = 13.6$ after including the Helium factor. The HCN conversion factor is also expected to decrease for U/LIRGs, but is much less constrained than the CO conversion factor \citep{Gao2004a}.

For normal disk galaxies, the commonly used values of $\alpha_{\text{HCN}} = 13.6$ and $\alpha_{\text{CO}} = 4.35$ have a ratio of $\alpha_{\text{HCN}}$/$\alpha_{\text{CO}} \approx 3.1$. Following the discussion in \cite{Bemis2023}, we apply a fixed ratio of $\alpha_{\text{HCN}}$/$\alpha_{\text{CO}} \approx 3.1$ for our U/LIRG sample, which combines with the U/LIRG value of $\alpha_{\text{CO}} = 1.1$ to give $\alpha_{\text{HCN}} \approx 3.5$ in U/LIRGs. The physical motivation for the decreased $\alpha_{\text{CO}}$ in more extreme systems like U/LIRGs is that CO is optically thick and the gas temperature, density, and line width are all higher \citep{Bolatto2013}. All these factors together mean that there is less mass of gas for a given CO luminosity. Similar physical arguments should apply to HCN emission, which is also thought to be optically thick (e.g., \citealt{Jimenez2017}) and higher temperatures, densities, and line widths are expected for HCN observations in U/LIRGs \citep{Solomon1992, Gao2004a, Gracia2006, Privon2015, Imanishi2019}.

Recently, \cite{Wilson2023} found a constant CN/HCN intensity ratio of $R_{\text{CN,HCN}} = I_{\text{CN}}/I_{\text{HCN}} = 0.86 \pm 0.07$ (standard deviation of 0.2) for the CN (1-0) and HCN (1-0) lines in a sample of 9 nearby star-forming galaxies, which included 4 U/LIRGs. The physical scales discussed in \cite{Wilson2023} are $30-400$ pc, roughly comparable with our $500$ pc resolution. We convert the CN intensity to HCN intensity using $R_{\text{CN,HCN}}$ and refer the reader to \cite{Wilson2023} for further discussion on the similarities of CN and HCN as dense gas tracers. Thus, we calculate $f_{\text{dense}}$ from the $I_{\text{CN}}/I_{\text{CO}}$ ratio using
\begin{equation}
    f_{\text{dense}} = \alpha_{\text{HCN}} / \alpha_{\text{CO}} \times 1 / R_{\text{CN,HCN}} \times I_{\text{CN}} / I_{\text{CO}},
    \label{eqn:fdense_obs}
\end{equation}
with $\alpha_{\text{HCN}}$/$\alpha_{\text{CO}} = 3.1$ and $R_{\text{CN,HCN}} = 0.86$, so that $f_{\text{dense}} = 3.6 \times I_{\text{CN}}/I_{\text{CO}}$.

\subsection{Star formation rates}
\label{subsection:sfr}

Since U/LIRGs contain large amounts of dust, infrared or radio emission must be used to obtain accurate SFRs. We calculate global SFRs using both global infrared luminosity and resolved radio continuum maps. The global infrared luminosities are taken from the original GOALS' publication \citep{Armus2009} and have been corrected for our choice of cosmology and distance. We calculate SFR from infrared luminosity using
\begin{equation}
    \text{log}(\text{SFR}) = \text{log}(L_{\text{IR}}) - \text{log}(C_{\text{IR}}),
    \label{eqn:sfr_LIR}
\end{equation}
where SFR is in M$_\odot$ yr$^{-1}$, $L_{\text{IR}}$ is the infrared luminosity in ergs s$^{-1}$, and log(C$_{\text{IR}}$) $= 43.41$ for $L_{\text{IR}}$ covering the wavelength range $3-1100 \ \mu\text{m}$ \citep{Murphy2011, Kennicutt2012}. The resulting SFRs span $\sim10-300$ $M_{\odot}$ yr$^{-1}$ (Table \ref{tab:measured_data}). We note that while 11 of 16 of our sources have evidence of an AGN (Table \ref{tab:sample_data}), we have not corrected the $L_{\text{IR}}$ emission for AGN contamination. Since AGN can contribute to the total infrared luminosity, we may overestimate the SFR in galaxies with AGN.

We use our resolved radio continuum images (Section \ref{subsec:alma}) to obtain the global radio continuum flux density for our 16 galaxies (Table \ref{tab:measured_data}). We calculate the $110$ GHz radio continuum luminosity as
\begin{equation}
    L_{\nu} = 4 \pi D_{\text{L}}^{2} (1 + z)^{-1} S_{\text{110 GHz}},
    \label{eqn:lum_radio}
\end{equation}
where $z$ is the redshift, $D_{\text{L}}$ is the luminosity distance, and $S_{\text{110 GHz}}$ is the observed flux density \citep{Solomon2005}. We convert from radio continuum to SFR using the standard thermal-only equation \citep{Murphy2011}
\begin{equation}
    \text{SFR} = 4.6\times10^{-28} \left(\frac{T_{\text{e}}}{10^{4} \ \text{K}}\right)^{0.45} \left(\frac{\nu}{\text{GHz}}\right)^{0.1} L_{\nu},
    \label{eqn:sfr_radio}
\end{equation}
where SFR has units of $M_{\odot}$ yr$^{-1}$ and $L_{\nu}$ has units of erg s$^{-1}$ Hz$^{-1}$. $T_{\text{e}}$ is the electron temperature, which we assume to be $10^{4}$ K, and for the continuum we use $\nu = 110$ GHz. The SFR calculated from the radio continuum spans $\sim10-650$ $M_{\odot}$ yr$^{-1}$ (Table \ref{tab:measured_data}).

The 110 GHz emission from dusty systems may be contaminated by synchrotron or dust emission. \cite{Wilson2019} estimate that dust will contribute $10-15\%$ to the radio continuum for NGC 3256. \cite{Sakamoto2017} found a contamination of $\sim40\%$ in the highly obscured western nucleus of Arp 220. It is difficult to quantify the specific dust contribution to our 110 GHz luminosities without spectral energy distribution fitting. We do not correct for any dust contamination, which may result in an overestimate of the SFRs calculated using the radio continuum.

Synchrotron emission is likely present in the galaxies in our sample which have strong AGN or Seyfert nuclei. We do not mask out the positions of the AGN in our global radio continuum flux density calculations as any AGN effect is blended with other radio continuum emission at our $500$ pc resolution. As with dust contamination, any AGN contribution to the radio continuum may result in an overestimate of our SFRs.

\begin{figure*}
    \captionsetup[subfigure]{labelformat=empty}
    \centering
    \begin{subfigure}{0.87\textwidth}
        \centering
        \includegraphics[width=\textwidth]{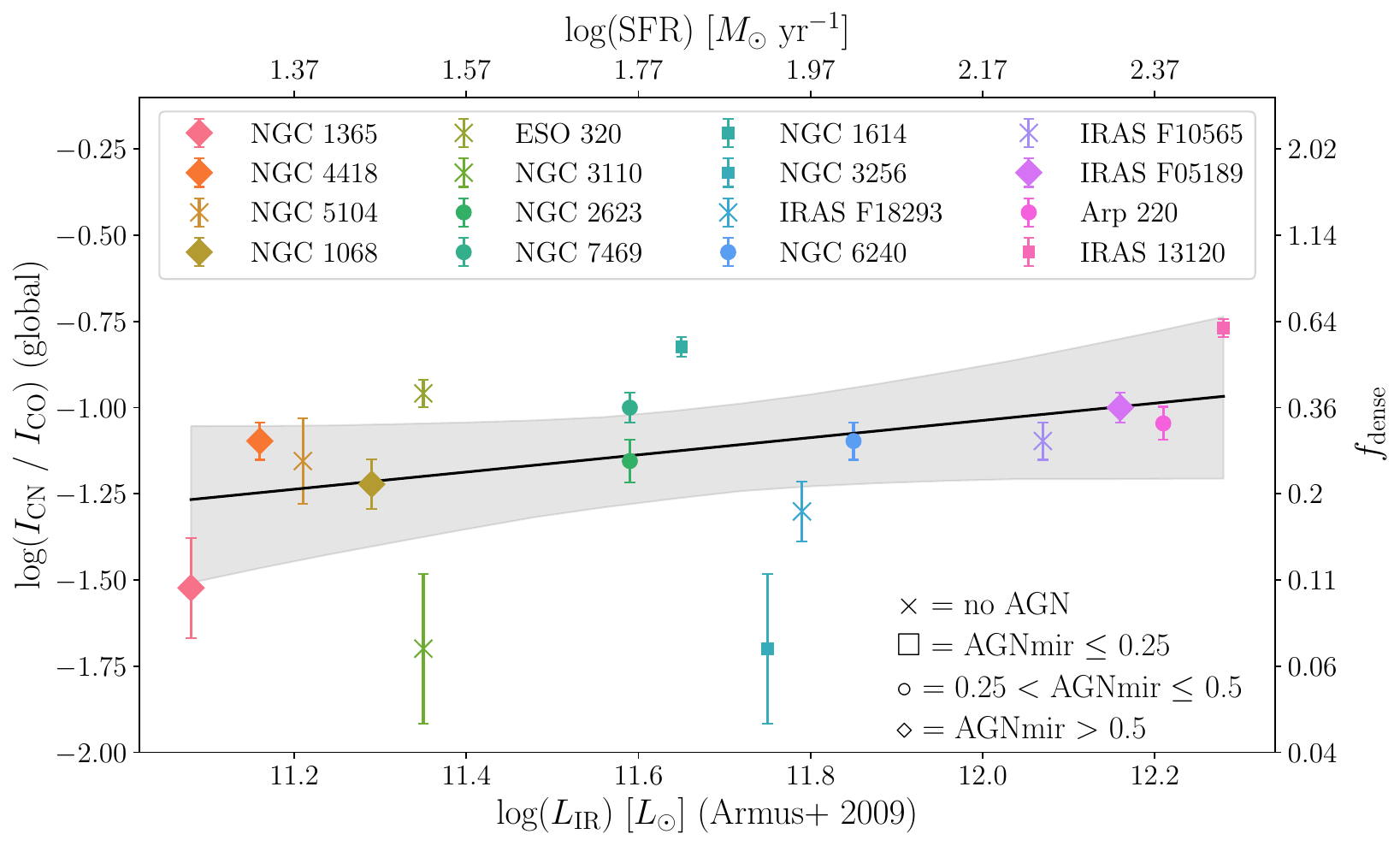}
        \caption{\textit{}}
        \label{fig:IR_vs_CN_CO}
    \end{subfigure}
    \begin{subfigure}{0.87\textwidth}
        \centering
        \includegraphics[width=\textwidth]{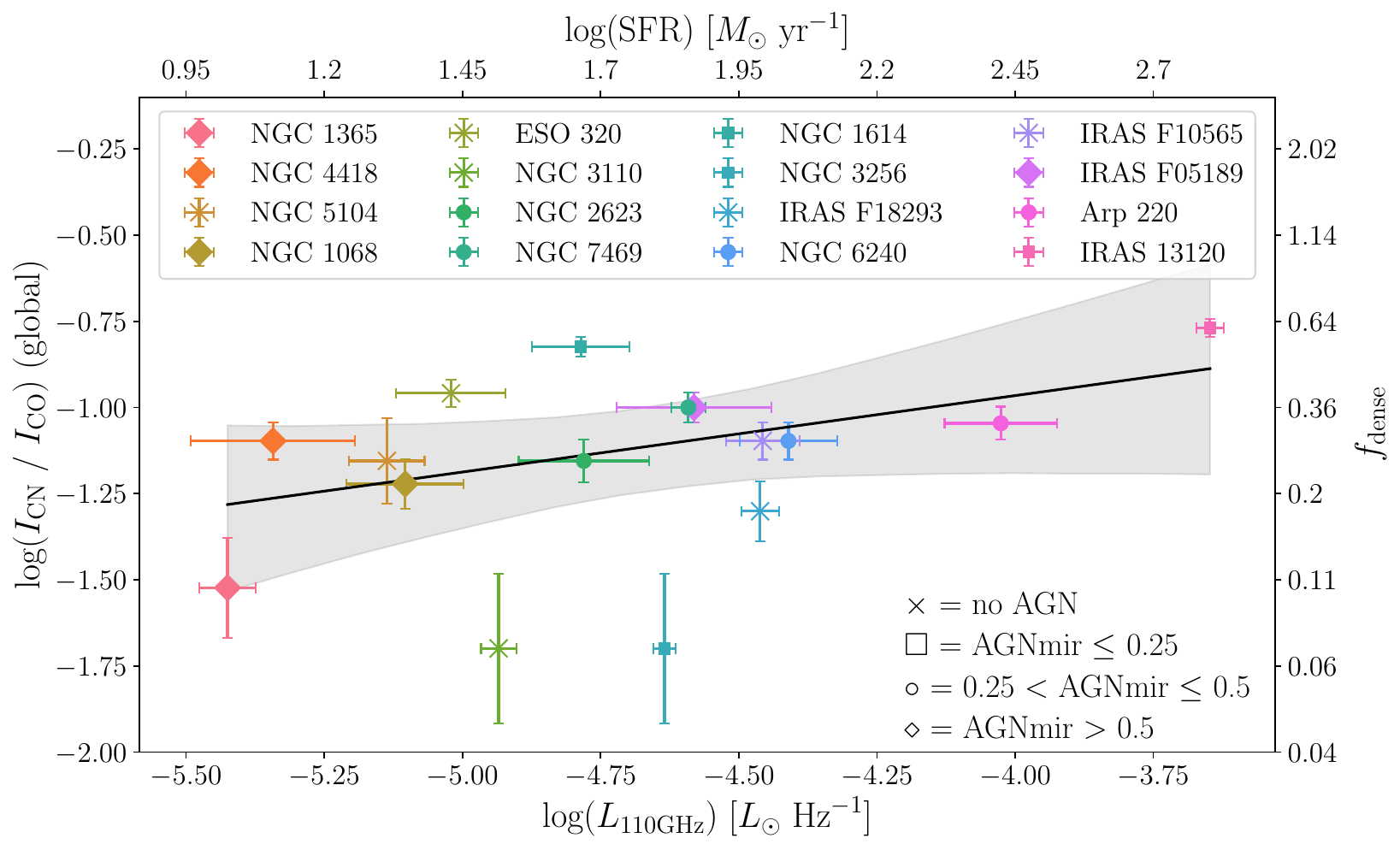}
        \caption{\textit{}}
        \label{fig:radio_vs_CN_CO}
    \end{subfigure}
    \caption{Global $I_{\text{CN}}/I_{\text{CO}}$ and dense gas fraction ($f_{\text{dense}}$) versus $L_{\text{IR}}$ (top), $L_{\text{110 GHz}}$ (bottom), and the inferred star formation rates. SFR has a significant correlation with $f_{\text{dense}}$ in U/LIRGs. Black solid lines show the best fit using \texttt{Linmix} with the 95\% confidence interval in the shaded region. Galaxies with AGN$_{\text{MIR}}<0.25$ are shown as square symbols, $0.25<$ AGN$_{\text{MIR}}<0.5$ as circle symbols, AGN$_{\text{MIR}}>0.5$ as diamond symbols \citep{DiazSantos2017}, and systems with no known AGN as crosses. The error bars shown for $L_{\text{110 GHz}}$ do not include a 5\% ALMA calibration uncertainty.}
    \label{fig:SFR_vs_CN_CO}
\end{figure*}

\begin{figure}
    \includegraphics[width=0.99\columnwidth]{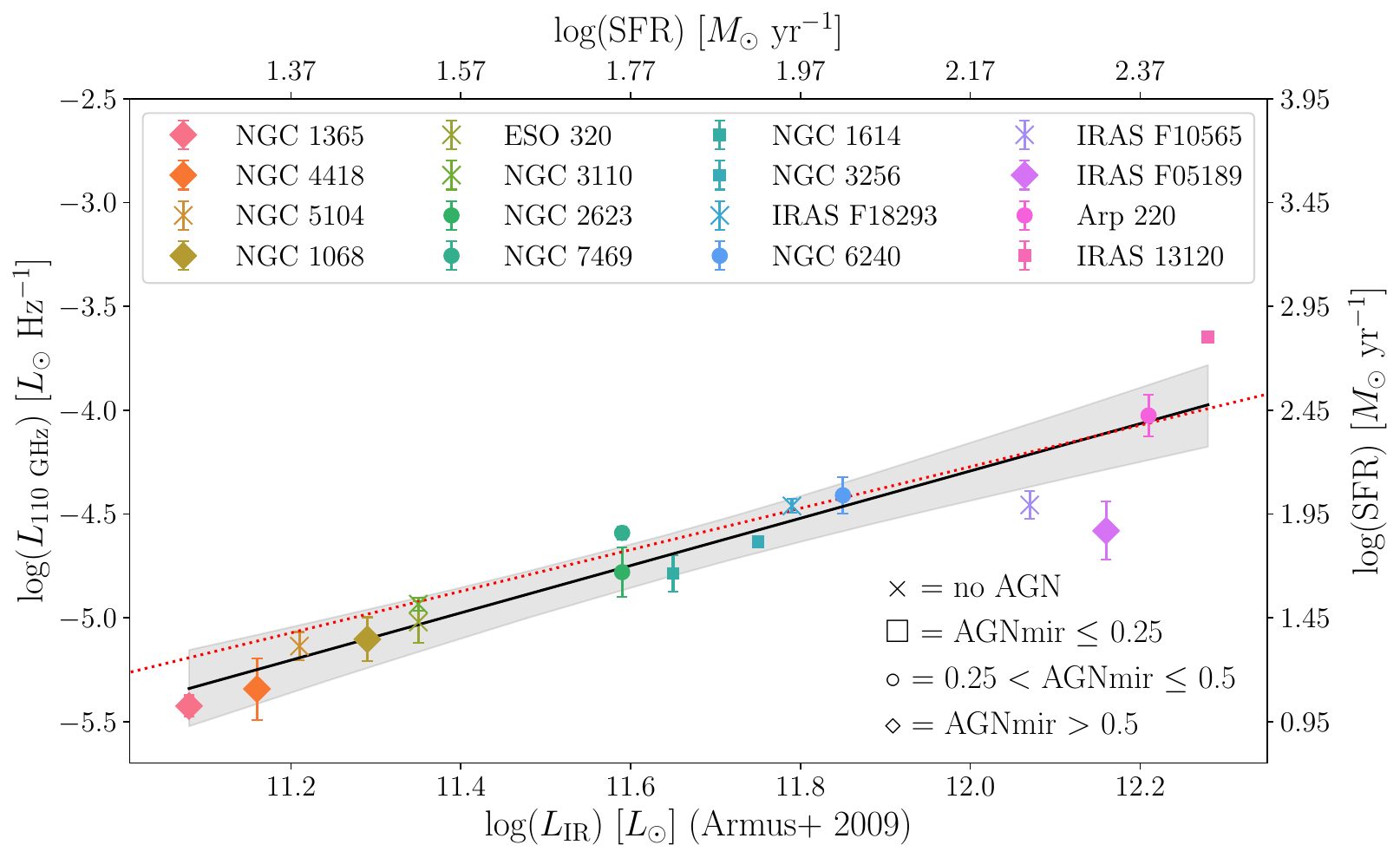}
    \caption{Global radio luminosity ($L_{\text{110 GHz}}$) versus global infrared luminosity ($L_{\text{IR}}$). Both axes show the corresponding star formation rates calculated using Equations \ref{eqn:sfr_LIR} and \ref{eqn:sfr_radio}. The red dotted line shows the 1-to-1 ratio between the star formation rates and the black solid line is the fit to the data. Symbols are as described in Figure \ref{fig:SFR_vs_CN_CO}. There is a strong correlation (Spearman R=$0.96$, p-value $\ll 0.01$) between these two star formation rate indicators.}
    \label{fig:SFR_comp}
\end{figure}

\begin{figure*}
    \captionsetup[subfigure]{labelformat=empty}
    \begin{subfigure}{0.55\textwidth}
        \centering
        \includegraphics[width=\textwidth]{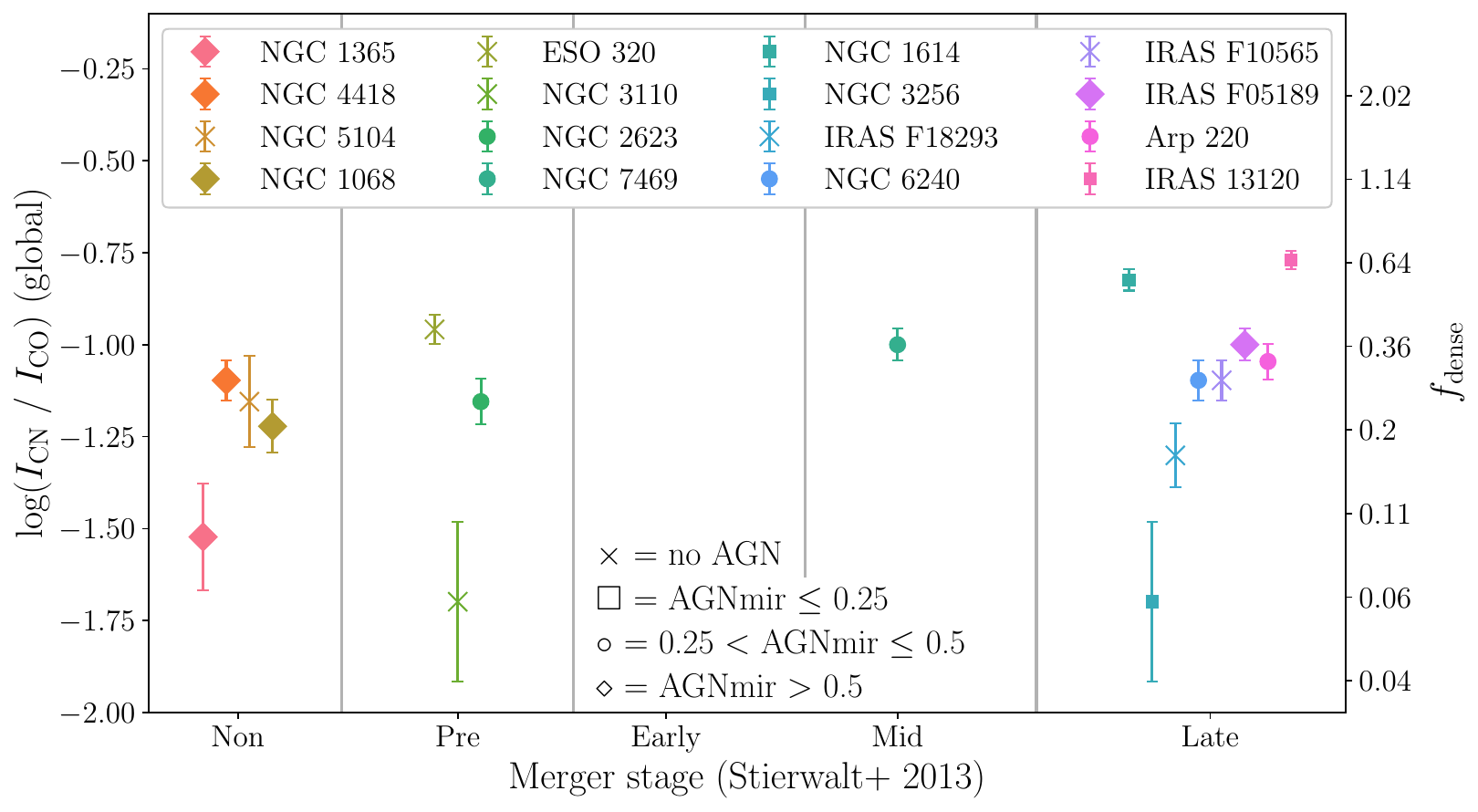}
        \caption{\textit{}}
        \label{fig:merger_physical}
    \end{subfigure}
    \hfill
    \begin{subfigure}{0.44\textwidth}
        \centering
        \includegraphics[width=\textwidth]{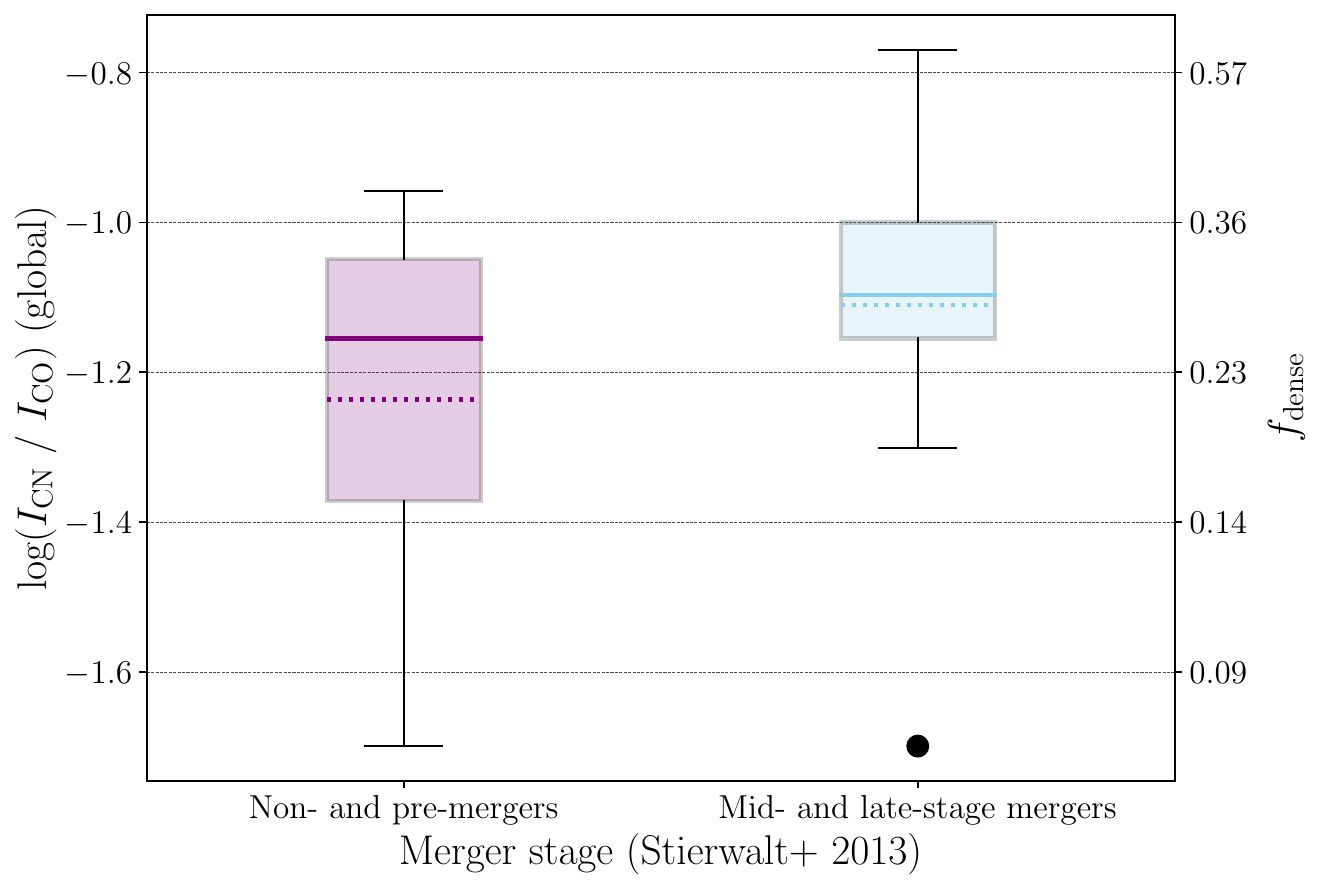}
        \caption{\textit{}}
        \label{fig:merger_physical_box}
    \end{subfigure}%
    \caption{\textit{Left}: Global $I_{\text{CN}}/I_{\text{CO}}$ ratios ($f_{\text{dense}}$) versus merger stage from \cite{Stierwalt2013}. There is no significant trend in $f_{\text{dense}}$ with merger stage. Symbols are as in Figure \ref{fig:SFR_vs_CN_CO}. \textit{Right}: Average global values show no statistical difference between $f_{\text{dense}}$ in ``early'' (purple, left) and ``late'' (blue, right) merger stages. The boxes extend from the first to third quartiles of the data and the whiskers extend from the minimum to the maximum of the data, with the black circle an outlier in the sample of mid- and late-stage mergers (NGC 3256). The solid and dotted horizontal lines give the median and mean of the data, respectively.}
    \label{fig:merger_vs_CN_CO}
\end{figure*}

\begin{figure*}
    \centering
    \captionsetup[subfigure]{labelformat=empty}
    \begin{subfigure}{0.99\textwidth}
        \centering
        \includegraphics[width=\textwidth]{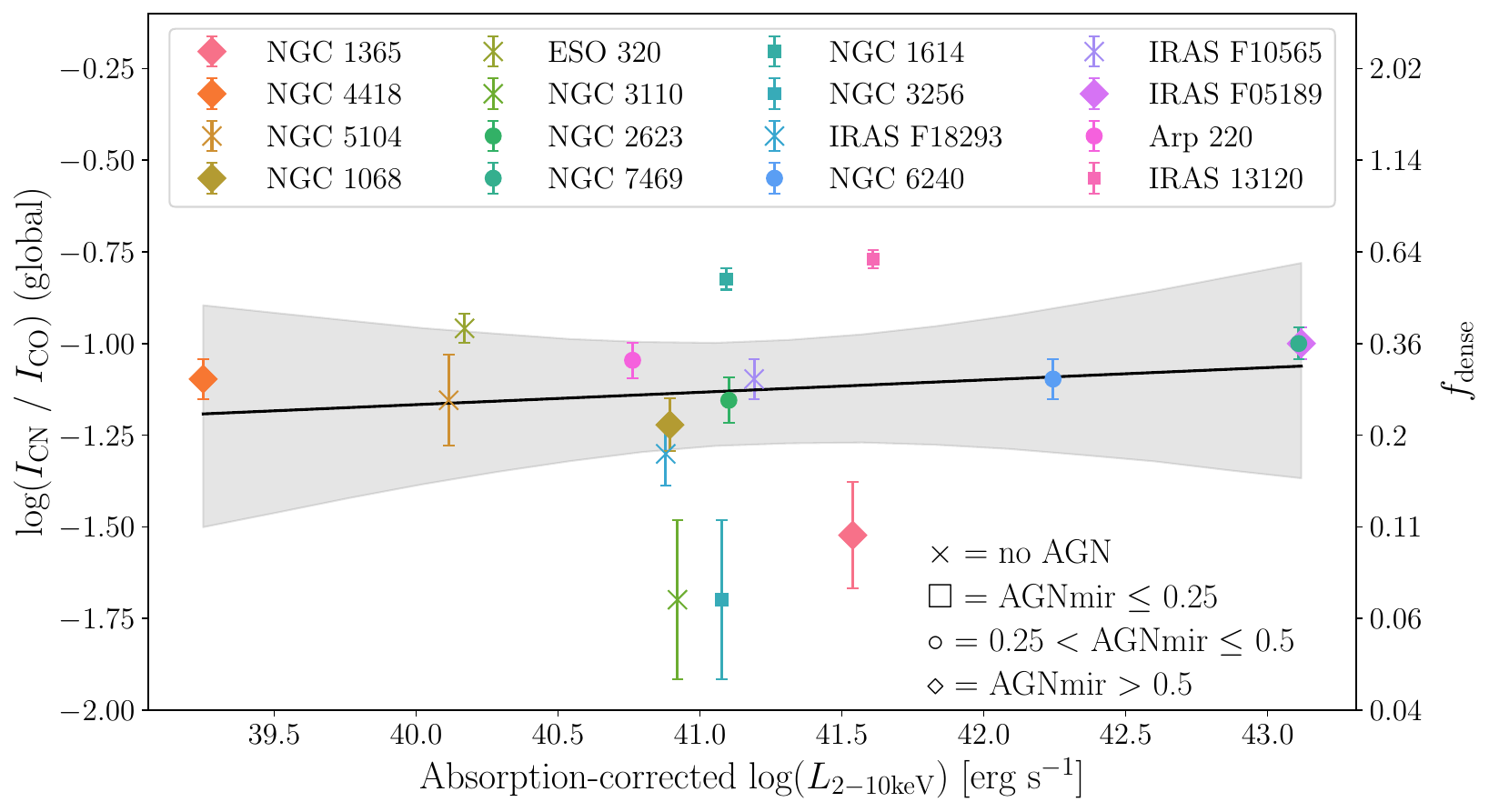}
        \caption{\textit{}}
        \label{fig:X_ray_vs_global_CN_CO}
    \end{subfigure}
    \begin{subfigure}{0.99\textwidth}
        \centering
        \includegraphics[width=\textwidth]{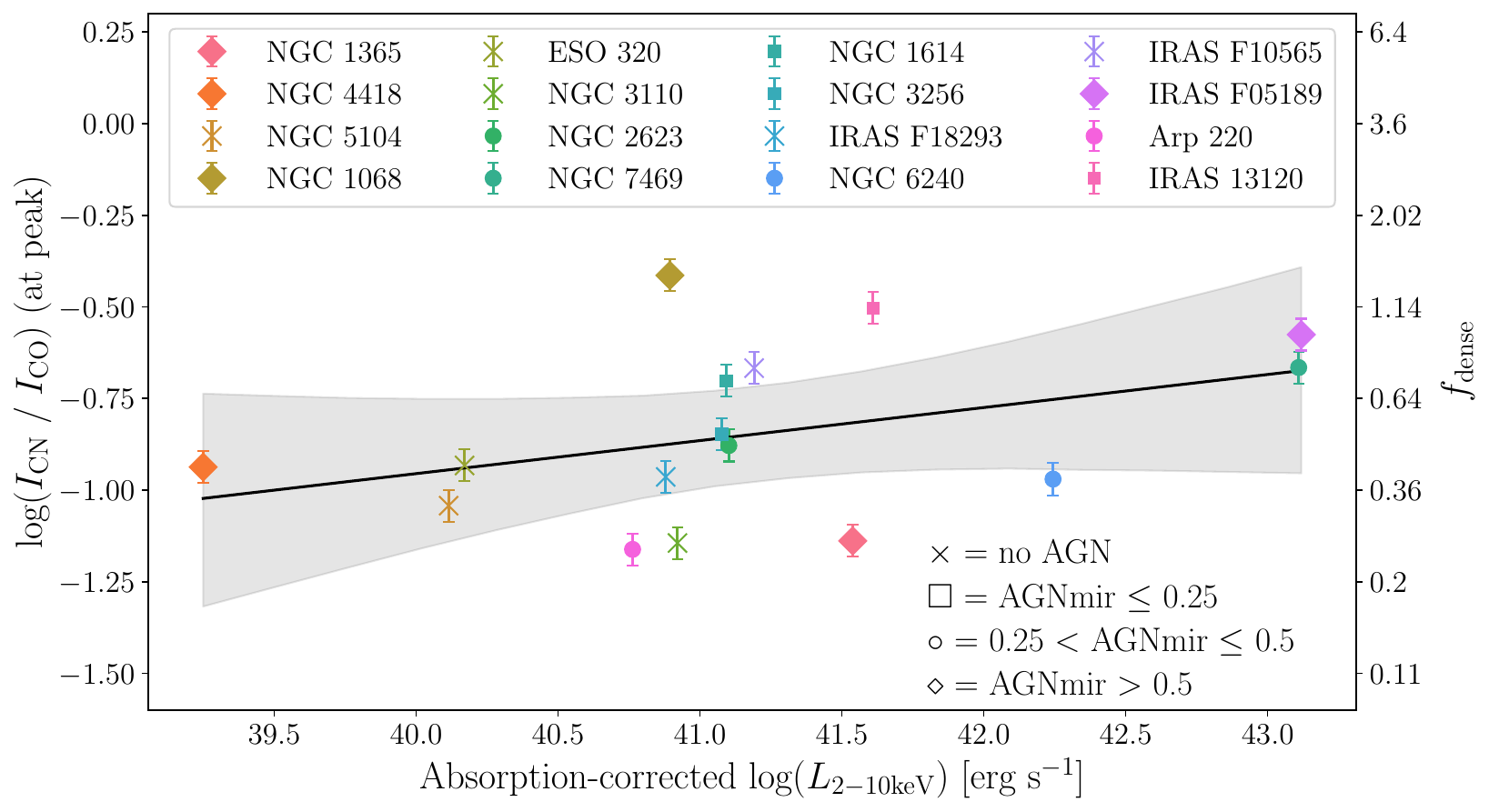}
        \caption{\textit{}}
        \label{fig:X_ray_vs_peak_CN_CO}
    \end{subfigure}%
    \caption{Global (top) and peak (bottom) $I_{\text{CN}}/I_{\text{CO}}$ ratios ($f_{\text{dense}}$) versus hard X-ray luminosity ($L_{\text{2-10 keV}}$), including Linmix fits to the data. Although there is no discernible trend between global dense gas fraction and hard X-ray luminosity, there is a statistically significant correlation between dense gas fraction at the peak X-ray pixel and the hard X-ray luminosity.}
    \label{fig:X_ray_vs_CN_CO}
\end{figure*}

\subsection{Compiled global properties}
\label{subsec:compiled}

The global properties that we have compiled to compare with the $I_{\text{CN}}/I_{\text{CO}}$ ratio are tabulated in Tables \ref{tab:measured_data} and \ref{tab:sample_data}. \cite{Stierwalt2013} assigned the merger stages via visual inspection of IRAC 3.6 $\mu$m images and mid-infrared spectra. They defined non-mergers (``N'') as galaxies with no merger activity or massive neighbours, pre-mergers (``a'') as galaxy pairs which have not had a first encounter, mid-stage mergers (``c'') as those showing signs of merger activity such as amorphous disk features and tidal tails, and late-stage mergers (``d'') as galaxies having a common envelope surrounding two nuclei. When applicable, \cite{Stierwalt2013} used higher resolution images and spectroscopic classifications from the literature to help with their classification of merger stages. Our sample consists of 4 non-mergers, 3 pre-mergers, 1 mid-stage merger, and 8 late-stage mergers (Table \ref{tab:sample_data}).

In Paper I, we broadly classify the type of AGN found in each galaxy as obscured, optically identified (Seyfert 1, Seyfert 2, or LINER), or with no evidence of an AGN. We list the contribution from the AGN to the mid-infrared luminosities of our galaxies (AGN$_{\text{MIR}}$) from \cite{DiazSantos2017} in Table \ref{tab:sample_data}. The Galactic absorption-corrected hard X-ray luminosities, $L_{\text{2-10 keV}}$, were compiled from various sources. 6 of our galaxies were observed with \textit{Chandra} by the GOALS team in C-GOALS I \citep{Iwasawa2011} and an additional 7 galaxies were observed in C-GOALS II \citep{Torres2018}. \cite{Iwasawa2011} and \cite{Torres2018} both report hard X-ray luminosities in the $2-7$ keV bands and extrapolate to the $2-10$ keV band. Of the remaining 3 galaxies in our sample, NGC 5104 has no conclusive evidence for an AGN \citep{Gonzalez2021}, but was detected by \textit{NUSTAR} in hard X-rays thought to originate from star formation \citep{Privon2020}. Hard X-rays were detected in ESO 320-G030 by \textit{XMM-Newton} \citep{Pereira2011} and this galaxy is also thought to not host an AGN \citep{Gonzalez2021}. NGC 7469 has a well-known Seyfert 2 AGN that has been studied in hard X-rays \citep{Blustin2003, Pereira2011, Liu2014, Mehdipour2018}. We use the absorption-corrected $2-10$ keV luminosity from \cite{Liu2014}, which combined data from \textit{XMM-Newton}, \textit{Chandra}, \textit{Suzaku}, and \textit{Swift}.

For galaxies with \textit{Chandra} data, the location of the hard X-ray peak is plotted as a black cross on the $I_{\text{CN}}/I_{\text{CO}}$ ratio maps in Figure \ref{fig:CN_CO_maps_w_xray}. This hard X-ray peak corresponds to star forming regions, AGN, outflows, or a combination of these sources. The X-ray peak roughly lines up with the peak CO intensity and also with the peak radio continuum emission. Therefore, we use the location of the hard X-ray peak for our analysis and comparison of the $I_{\text{CN}}/I_{\text{CO}}$ ratio at the ``peak'' (Table \ref{tab:measured_data}).

NGC 4418 and NGC 6240 have double X-ray peaks. In NGC 6240, these two X-ray peaks correspond to dual Compton-Thick AGN \citep{Iwasawa2011}. In NGC 4418, the galaxy has a single, heavily obscured AGN and the two X-ray peaks are likely two sides of the obscured AGN emission, with the true X-ray source located between these two peaks \citep{Torres2018}. For the galaxies with two X-ray peaks, we choose the X-ray peak which is closest to the peak CO intensity. For the 3 galaxies which have no \textit{Chandra} data, we use the pixel corresponding to the peak CO intensity (shown as a gray plus sign in Figure \ref{fig:CN_CO_maps_w_xray}). 

\section{Results}
\label{sec:results}

In this paper, we use the CN/CO intensity ratio to estimate the dense gas fraction. Due to the novelty of this method, most of our comparison with previous work on the dense gas fraction will use HCN-based literature results. Although CN/CO and HCN/CO trace each other well in star-forming galaxies \citep{Wilson2023}, we caution the reader to keep in mind some of the possible physical and chemical differences between HCN and CN, such as differing optical depths, when interpreting our results. Our analysis relies on assumptions for the conversion factors of both CO and HCN in calculating $f_{\text{dense}}$ and for converting the CN emission to HCN. We use both Spearman and Pearson rank coefficients to test the strength and significance of correlations between $I_{\text{CN}}/I_{\text{CO}}$ and other quantities. We find three significant correlations (p-values $<0.1$ in one or both of the rank tests): global $I_{\text{CN}}/I_{\text{CO}}$ and $L_{\text{IR}}$; global $I_{\text{CN}}/I_{\text{CO}}$ and $L_{\text{110 GHz}}$; $I_{\text{CN}}/I_{\text{CO}}$ at the peak pixel and $L_{\text{2-10 keV}}$.

\subsection{Dense gas fraction and star formation rate}
\label{subsubsec:dense_and_SFR}

In Figure \ref{fig:SFR_vs_CN_CO}, we compare the observed $I_{\text{CN}}/I_{\text{CO}}$ ratio to $L_{\text{IR}}$ (top) and $L_{\text{110 GHz}}$ (bottom). We also include the conversion to the physical quantities $f_{\text{dense}}$ and SFR. Both plots show positive correlations, although with significant scatter ($L_{\text{IR}}$: p-value $<0.08$, Pearson R $=0.45$; $L_{\text{110 GHz}}$: p-value $<0.05$, Pearson R $=0.49$). Although these are global $I_{\text{CN}}/I_{\text{CO}}$ values, the scatter is likely a direct result of the internal variations observed in the resolved $I_{\text{CN}}/I_{\text{CO}}$ ratio for the U/LIRG sample (see Figure \ref{fig:CN_CO_maps_w_xray}), which has been averaged out in our global measurement \citep{Ledger2024}. Figure \ref{fig:SFR_comp} compares the two SFR tracers; they are strongly correlated with a Spearman rank correlation coefficient of $0.96$ and a p-value of $\ll0.01$. This result agrees with previous observations that global radio continuum and IR emission are correlated for various galaxy luminosities and types \citep{Liu2010}, including U/LIRGs \citep{Yun2001}.

Comparing $I_{\text{CN}}/I_{\text{CO}}$ with $L_{\text{IR}}$ and $L_{\text{110 GHz}}$ probes the physical connection between $f_{\text{dense}}$ and SFR. A significant, positive correlation between $f_{\text{dense}}$ and SFR is consistent with previous work which has primarily used HCN as the tracer of dense gas. \cite{Aalto1995} did not find a correlation between $f_{\text{dense}}$ (via HCN/CO) and far-infrared emission in a sample of 11 interacting galaxies and mergers, 5 of which were LIRGs and the rest normal, star-forming galaxies. This lack of correlation may result from the diverse nature of their galaxy sample. \cite{Gao2004a} compared $f_{\text{dense}}$ (via HCN/CO) and SFR (via $L_{\text{IR}}$) and found a weak correlation in their sample of U/LIRGs, but no correlation on the lower luminosity end in more normal disk galaxies. All galaxies above a certain $f_{\text{dense}}$ were found to be U/LIRGs with high SFRs, but not all U/LIRGs had a high $f_{\text{dense}}$, which is in agreement with our Figure \ref{fig:SFR_vs_CN_CO}. Their conclusion was that a high value of $f_{\text{dense}}$ is a good indicator of a starbursting galaxy.

Additionally, \cite{Gracia2006} and \cite{Juneau2009} found a positive correlation between $f_{\text{dense}}$ (via HCN/CO) and $L_{\text{IR}}$ with quite a large scatter in the relation, particularly toward the U/LIRG end. \cite{Juneau2009} compared observations and simulations of $f_{\text{dense}}$ (based on \citealt{Narayanan2008}) and found that both show little to no correlation with lower $L_{\text{IR}}$ values and then an increase in $f_{\text{dense}}$ for higher $L_{\text{IR}}$. The authors suggest that the galaxies with a larger reservoir of dense gas (and/or a higher dense gas fraction) have more capability to fuel powerful starbursts and AGN activity, both of which are associated with higher IR luminosities. In a sample of roughly 20 merger remnants, \cite{Ueda2021} found a weak correlation between $f_{\text{dense}}$ (via HCN/CO) and SFR ($L_{\text{IR}}$), particularly towards the higher end of $L_{\text{IR}}$ values. The presence of a weak correlation and a large scatter in $f_{\text{dense}}$ with SFR persists in galaxies at higher redshift ($z>1$, \citealt{Gowardhan2017, Oteo2017}).

The observed correlation between $f_{\text{dense}}$ (measured using CN/CO) and SFR (using both radio and infrared emission) in our sample confirms that U/LIRGs with higher IR luminosities and SFRs tend to have higher fractions of dense gas, on average. The scatter in the correlation is likely driven by physical differences between galaxies resulting from, for example, morphological differences, compact starburst regions, and AGN.

\subsection{Dense gas fraction and merger stage}
\label{subsubsec:dense_and_merger}

We plot the global $I_{\text{CN}}/I_{\text{CO}}$ ratio ($f_{\text{dense}}$) versus merger stage in Figure \ref{fig:merger_vs_CN_CO}. We also bin the data into ``early'' and ``late'' mergers, where ``early'' mergers are both non- and pre-mergers, while ``late'' mergers are mid- and late-stage mergers\footnote{Note that we have no ``early mergers'', as originally classifed as type `b' in \cite{Stierwalt2013}, in our sample.}. The mean $I_{\text{CN}}/I_{\text{CO}}$ for ``early'' and ``late'' mergers are $0.07\pm0.01$ and $0.09\pm0.02$, respectively. Due to our small sample sizes, we use an Anderson-Darling (AD) test to compare the $I_{\text{CN}} / I_{\text{CO}}$ ratio between the two samples \citep{Scholz1987}. The AD test shows that there is no statistical difference (p-value $>0.8$) in the global $I_{\text{CN}}/I_{\text{CO}}$ between ``early'' and ``late'' stage mergers.

Simulations of mergers allow for a more direct comparison of how the gas density distribution changes during the merger process. While galaxy simulations suggest that $f_{\text{dense}}$ will increase in a galaxy during the merging process \citep{Juneau2009, Moreno2019}, we find no connection between $f_{\text{dense}}$ and merger stage in our sample of U/LIRGs. Since our sample only has a few galaxies in each merger stage bin, any real physical connection between an average dense gas fraction and merger stage may be hidden by individual galaxy differences. Observations of a larger sample of galaxies covering a wide range of merger stages are needed to resolve the discrepancy between simulations and observations.

Although some simulations show that interactions and merger events increase the amount of dense gas in the central kiloparsec region of galaxies (e.g., \citealt{Cenci2024}), we find no connection between $f_{\text{dense}}$ and merger stage at the X-ray peak in our sample of U/LIRGs. The mean $I_{\text{CN}}/I_{\text{CO}}$ ratio at the location of the peak X-ray emission for ``early'' and ``late'' mergers is $0.15\pm0.04$ and $0.17\pm0.03$, respectively; both values are larger than the mean global values. The results of an AD test suggest that there is no statistical difference (p-value $>0.5$) in the $I_{\text{CN}}/I_{\text{CO}}$ at the peak pixel between ``early'' and ``late'' stage mergers. Once again, resolved observations in the centres of a statistically significant sample of galaxies in different merger stages are required to resolve the discrepancy between observations and simulations. Furthermore, changes in line excitation created by e.g. central starbursts and/or AGN can further complicate the measurement of central dense gas fraction as a function of merger stage.

\subsection{Dense gas fraction and hard X-ray luminosity}
\label{subsubsec:dense_and_xray}

We compare the global $I_{\text{CN}}/I_{\text{CO}}$ ratio with the global hard X-ray luminosity ($L_{\text{2-10 keV}}$) in Figure \ref{fig:X_ray_vs_CN_CO} (top). For galaxies with a confirmed AGN (see Table \ref{tab:sample_data}), the hard X-ray luminosity is likely to be strongly dominated by the AGN. The correlation between global $I_{\text{CN}}/I_{\text{CO}}$ and $L_{\text{2-10 keV}}$ is not statistically significant (p-value $>0.37$). Thus, we find no connection between $f_{\text{dense}}$ and hard X-ray luminosities on global scales.

The extent of X-ray emission and impact on gas properties is much more localized than, for example, star formation (e.g., \citealt{Wolfire2022}). Resolved observations of nearby star-forming galaxies and U/LIRGs (including many overlapping with our sample) have also found multiple different dense gas tracers have increased emission in the centres relative to the disks (e.g., \citealt{Meier2014, Xu2015, Privon2017, Li2024}). Therefore, it is useful to explore the dense gas properties at the peak X-ray position rather than a global average. For many of our galaxies, the strong X-ray emission at the peak pixel is a result of concentrated X-ray emission and the ``global'' X-ray is practically the same as a ``peak'' X-ray. We find in all galaxies that $f_{\text{dense}}$ increases at the peak X-ray pixel (near the galaxy nuclei) compared to the global value (except Arp 220, see Paper I).

We compare $I_{\text{CN}}/I_{\text{CO}}$ at the peak X-ray pixel with the global $L_{\text{2-10 keV}}$ (Figure \ref{fig:X_ray_vs_CN_CO}, bottom) and find a significant correlation between the two quantities (p-value $<0.09$, Spearman R$=0.44$), unlike the global $I_{\text{CN}}/I_{\text{CO}}$ comparison. We interpret this difference as a co-localization of higher $f_{\text{dense}}$ and the source of the X-ray emission.

Previous work comparing dense gas and X-ray properties has commonly used the HCN line. In the presence of strong X-ray radiation fields, such as near AGN, gas temperatures may become high enough (more than a few hundred Kelvin) to make the chemical pathways which convert CN into HCN become important (e.g., \citealt{Harada2013}). However, at our resolution of $500$ pc we are likely not sufficiently resolving any XDRs where these reactions will become important, and therefore can compare our CN-based results with those in the literature which have used HCN as the dense gas tracer.

\cite{Izumi2016} observed a positive correlation between HCN (1-0) emission ($\sim200$ pc scales) and $L_{\text{2-10 keV}}$ in 10 Seyfert galaxies and interpreted it as a correlation between dense gas mass and black hole accretion rate. The authors argued that the dense gas in the CNDs of the galaxies was helping to feed accretion onto the supermassive black holes. In contrast, using $20-50$ keV hard X-rays, \cite{Kawamuro2021} observed lower $f_{\text{dense}}$ with increasing X-ray luminosity in the central region of a sample of $26$ galaxies at $100-600$ pc resolution. We are likely not resolving any individual CNDs in our sample, but are instead seeing an unresolved increase in $f_{\text{dense}}$ in the nuclear regions of our galaxies.

With higher resolution observations ($\sim10$ pc) of the Circinus galaxy, which hosts a strong Seyfert nuclei, \cite{Kawamuro2019} were able to find a spatial anti-correlation between dense gas and X-ray irradiated gas traced through highly ionized metal lines and argued X-rays were impacting the gas properties. In contrast, \cite{Garcia2010} found that the hard X-ray peak in \textit{Chandra} data coincides with the position of their observed peak CN (2-1)/CO(1-0) intensity ratio in NGC 1068. They also found a positive correlation between their resolved CN/CO intensity ratio and hard X-ray luminosity, which they argue is a result of X-ray gas chemistry and an X-ray dominated region in this galaxy. We are not able to resolve the scales which would allow us to see the destructive impact that hard X-ray emission may have on molecular gas. Therefore, our analysis averages together the effect of any negative AGN feedback on the molecular gas and the existence of more dense gas in the centres of these galaxies.

For 2 galaxies in our sample, IRAS 13120 and NGC 1068, $f_{\text{dense}}>1$ at the X-ray peak. $f_{\text{dense}}>1$ means that there is more dense gas than there is total molecular gas, which is not physically possible. IRAS 13120 is the most luminous ULIRG in our sample, with the highest SFR of a few hundred $M_{\odot}$ yr$^{-1}$, and NGC 1068 has a powerful Seyfert 2 nucleus. It is likely in these two systems that additional physical or chemical conditions in the central regions near the X-ray peak pixel are increasing the CN emission relative to CO, e.g., through CN enhancement or CO destruction (c.f. \citealt{Saito2022b}). A similar strong enhancement of CN relative to CO was observed in the outflow of Mrk 231 \citep{Cicone2020}. In our simple calculation of $f_{\text{dense}}$, we do not account for the unique excitation conditions and therefore we overestimate the peak $f_{\text{dense}}$ in these galaxies. 

\section{Conclusions}
\label{sec:conc}

We use the $I_{\text{CN}}/I_{\text{CO}}$ intensity ratio to estimate the dense gas fraction in a sample of 16 U/LIRGs and compare it to global galaxy properties such as SFR, merger stage, and hard X-ray luminosity. The correlations which are most statistically significant (p-values $<0.1$) are global $f_{\text{dense}}$ with SFR and $f_{\text{dense}}$ at the X-ray peak pixel with hard X-ray luminosity. Our main conclusions are the following.

\begin{enumerate}
    \item Using both infrared luminosity and radio continuum as SFR tracers, we find a significant correlation between global $f_{\text{dense}}$ and SFR in our sample of U/LIRGs.
    \item We find no correlation between $f_{\text{dense}}$ and merger stage for global or peak values of $f_{\text{dense}}$. This result agrees with previous observations showing little to no connection between dense gas and merger stage, but is in conflict with simulations \citep{Juneau2009, Moreno2019, Cenci2024}.
    \item We find a significant positive correlation between $f_{\text{dense}}$ at the peak X-ray pixel and hard X-ray luminosity. This result suggests that dense gas and hard X-ray emission are co-localized such that denser gas can foster AGN activation and growth (e.g., \citealt{Izumi2016}).
\end{enumerate}

\noindent Future work with this sample will compare $f_{\text{dense}}$ with resolved radio continuum continuum and star formation rate maps (Klimi et al. \textit{in prep.}). This work will allow us to remove sites of AGN contamination and compare with previous resolved works in normal star-forming galaxies (e.g., \citealt{Gallagher2018}).\newline

%


This paper makes use of the following ALMA data:

\noindent{ADS/JAO.ALMA\#2012.1.00306.S,} \\
ADS/JAO.ALMA\#2012.1.00657.S, \\
ADS/JAO.ALMA\#2013.1.00218.S, \\
ADS/JAO.ALMA\#2013.1.00991.S, \\
ADS/JAO.ALMA\#2013.1.01172.S, \\
ADS/JAO.ALMA\#2015.1.00003.S, \\
ADS/JAO.ALMA\#2015.1.00167.S, \\
ADS/JAO.ALMA\#2015.1.00287.S, \\
ADS/JAO.ALMA\#2015.1.01135.S, \\
ADS/JAO.ALMA\#2015.1.01191.S, \\
ADS/JAO.ALMA\#2016.1.00177.S, \\
ADS/JAO.ALMA\#2016.1.00263.S, \\
ADS/JAO.ALMA\#2017.1.00078.S, \\
ADS/JAO.ALMA\#2018.1.00223.S, \\
ADS/JAO.ALMA\#2018.1.01684.S, and \\
ADS/JAO.ALMA\#2019.1.01664.S.

ALMA is a partnership of ESO (representing its member states), NSF (USA), and NINS (Japan), together with NRC (Canada), MOST and ASIAA (Taiwan), and KASI (Republic of Korea), in cooperation with the Republic of Chile. The Joint ALMA Observatory is operated by ESO, AUI/NRAO, and NAOJ. The National Radio Astronomy Observatory is a facility of the National Science Foundation operated under cooperative agreement by Associated Universities, Inc. Any ALMA specification details in this research has made use of G. Privon et al. 2022, ALMA Cycle 9 Proposer’s Guide, ALMA Doc. 9.2 v1.4\footnote{https://almascience.eso.org/documents-and-tools/cycle9/alma-proposers-guide}.

This research has made use of the NASA/IPAC Extragalactic Database (NED) which is operated by the Jet Propulsion Laboratory, California Institute of Technology, under contract with the National Aeronautics and Space Administration. The computing resources available at NAOJ were essential for Paper I of this project. Data analysis was in part carried out on the Multi-wavelength Data Analysis System (MDAS) operated by the Astronomy Data Center (ADC), National Astronomical Observatory of Japan.

BL would like to thank Drs. Daisuke Iono and Toshiki Saito for hosting him at NAOJ and their guidance during the preparation of Paper I. BL acknowledges support from an NSERC Canada Graduate Scholarship-Doctoral. CDW acknowledges financial support from the Canada Council for the Arts through a Killam Research Fellowship. The research of CDW is supported by grants from the Natural Sciences and Engineering Research Council of Canada (NSERC) and the Canada Research Chairs program. NTA acknowledges support from NASA grants 80NSSC23K1611 and 80NSSC23K0484. TS acknowledges support from Japan Foundation for Promotion of Astronomy.

\facilities{ALMA, Chandra, IRAS, NUSTAR, Spitzer, Suzaku, Swift, and XMM-Newton.}


\software{This research has made use of the following software packages: \texttt{ASTROPY}\footnote{https://www.astropy.org/}, a community-developed core \texttt{PYTHON} package for astronomy \citep{astropy2013, astropy2018, astropy2022}, \textsc{CASA}\footnote{https://casa.nrao.edu/} \citep{McMullin2007}, \texttt{MATPLOTLIB}\footnote{https://matplotlib.org/} \citep{Hunter2007}, \texttt{NUMPY}\footnote{https://numpy.org/} \citep{harris2020}, RStudio\footnote{https://posit.co/download/rstudio-desktop/} \citep{R2023}, \texttt{SciPy}\footnote{https://scipy.org/} \citep{scipy2020}, and \texttt{SPECTRAL-CUBE}\footnote{https://spectral-cube.readthedocs.io/en/latest/} \citep{Ginsburg2019}.}





\bibliography{CO_CN_paper_2}{}
\bibliographystyle{aasjournal}



\end{document}